\documentclass[final,5p,times,twocolumn,number]{elsarticle}

\usepackage{amsmath, amssymb, geometry}
\allowdisplaybreaks
\usepackage{amssymb}
\usepackage{graphicx}
\usepackage{xcolor}
\usepackage[colorlinks]{hyperref}
\usepackage{orcidlink}
\usepackage{multirow}
\usepackage{subcaption}
\usepackage{makecell} 
\usepackage{cuted}

\usepackage{amssymb}
\usepackage{lipsum}

\journal{Physics Letters B}

\begin{document}

\begin{frontmatter}



\title{Dirac fields in LRS III spacetimes: a dynamical systems analysis}

\author[first]{Fabrizio Esposito \orcidlink{0000-0001-6883-152X}}
\ead{fabrizio.esposito01@edu.unige.it}
\author[first,second,third,fourth]{Sante Carloni \orcidlink{0000-0003-2373-2653}}
\ead{sante.carloni@unige.it}
\author[first,second,third]{Stefano Vignolo \orcidlink{0000-0003-2926-0650}}
\ead{stefano.vignolo@unige.it}
\affiliation[first]{organization={DIME,      Università di Genova},
            addressline={Via all'Opera Pia 15}, 
            city={Genova},
            postcode={16145}, 
            state={Italy},
            country={}}
\affiliation[second]{organization={INFN Sezione di Genova},
            addressline={Via Dodecaneso 33}, 
            city={Genova},
            postcode={16146}, 
            state={Italy},
            country={}}
\affiliation[third]{organization={GNFM, Istituto Nazionale di Alta Matematica},
            addressline={ P.le Aldo Moro 5}, 
            city={Roma},
            postcode={00185}, 
            state={Italy},
            country={}}
\affiliation[fourth]{INAF - Osservatorio Astronomico di Roma, via Frascati 33,
00040 - Monte Porzio Catone (Roma), Italy}

\begin{abstract}
Within the $1+1+2$ covariant formalism, we investigate Locally Rotationally Symmetric (LRS) class III spacetimes sourced by a self-gravitating Dirac field expressed in polar form. By recasting the covariant field equations into an autonomous dynamical system, we perform a phase-space analysis of their solutions. We show that the cosmological evolution asymptotically converges to a contracting Bianchi I spacetime, a result also supported by numerical integrations of the system.
\end{abstract}




\end{frontmatter}




\section{Introduction}\label{sec:introduction}

Building on Ehlers' pioneering work on relativistic hydrodynamics \cite{Ehlers1961}, Ellis and collaborators developed the covariant approach to General Relativity \cite{PhysRevD.40.1804,Ellis:1989ju,PhysRevD.42.1035,vanElst:1995eg,Ellis:1998ct}. Among its various formulations, the $1+1+2$ decomposition is particularly well suited to the study of Locally Rotationally Symmetric (LRS) spacetimes. Over the years, this formalism has proved to be a powerful framework for investigating a broad range of problems in cosmology and astrophysics \cite{Clarkson:2002jz,Clarkson:2007yp, Betschart:2004uu, Carloni:2017rpu, Carloni:2017bck, Naidu:2021nwh, Naidu:2022igk, Luz:2024yjm, Luz:2024xnd, Luz:2024lgi, Carloni:2006gy}. However, its application to Dirac fields is far from straightforward because of the presence of Clifford matrices and the nontrivial structure of the spinor covariant derivative. A natural way to overcome these difficulties is provided by the polar formalism for spinor fields \cite{Fabbri:2020ypd,Fabbri:2023dgv,Fabbri_2024,Fabbri_2025}. In this representation, the spinor field is parametrized in terms of a scalar density, a timelike velocity vector, a spacelike spin vector, and a chiral angle, thereby eliminating the explicit dependence on tetrads and gamma matrices. Moreover, the spinor covariant derivative becomes proportional to the spinor itself, allowing one to interpret the Dirac field as an effective relativistic fluid whose energy-momentum tensor can be treated within the standard covariant formalism. 

The first covariant formulation of self-gravitating Dirac fields within the $1+1+2$ framework was presented in  \cite{Vignolo:2025ppo}, where the velocity and spin vectors were identified with the preferred congruences defining the decomposition. Under this assumption, LRS I and LRS II spacetimes were shown to admit Dirac sources, whereas LRS III geometries were found to be incompatible with them. More recently \cite{Vignolo:2026qfn}, it was found that considering a generic frame coplanar with the spinor velocity-spin pair, solutions could be found also in LRS III spacetimes.

The present work is devoted to the dynamical investigation of this LRS III configuration. Since the corresponding Einstein-Dirac equations constitute a highly nonlinear system for which analytical solutions are not available, we employ the Dynamical Systems Approach (DSA) \cite{perko2012differential,Bahamonde:2017ize} to determine their global asymptotic behavior. After introducing suitable dimensionless variables normalized with respect to a combination of the expansion scalar and the spinor density, we recast the evolution equations into an autonomous system. We then identify its critical points, analyze their stability, and determine the asymptotic cosmological evolution. We show that all solutions evolve towards a contracting Bianchi I spacetime. The analytical results are further corroborated by direct numerical integrations of the complete dynamical system.

The paper is organized as follows. In Section \ref{sec:1+1+2} we briefly review the $1+1+2$ covariant formalism for LRS spacetimes. Section \ref{sec:polar} summarizes the polar decomposition of the Dirac field, while Section \ref{sec:LRS III} presents the covariant equations governing LRS III spacetimes sourced by a non-perfect spinor fluid. The dynamical systems analysis, together with the fixed points, their stability, and representative numerical solutions, is presented in Section \ref{sec:DSA}. Finally, Section \ref{sec:conclusions} contains our concluding remarks.

Throughout the paper, we adopt natural units, $(c=8\pi G = \hbar = 1)$, and the metric signature $(+,-,-,-)$.

\section{The \texorpdfstring{$1+1+2$}{} covariant formalism for LRS spacetimes}\label{sec:1+1+2}

In this work, we deal with LRS spacetimes, and for the reader's convenience, we present preliminary generalities of the $1+1+2$ decomposition of this class of spacetimes. 

To this end, let $v^{i}$ and $e^{i}$ be the timelike and spacelike unit vector fields that are respectively tangent to the two congruences which induce the $1+1+2$ splitting. They satisfy the relations
\begin{equation}
v^{i}v_{i}=1,\qquad e^{i}e_{i}=-1,\qquad e^{i}v_{i}=0.
\end{equation}
and can be associated with a class of observers with velocity $v^{i}$  that have chosen the integral curves of $e^{i}$ as reference spatial direction.

At each point of spacetime, the tangent space can be decomposed into the direct sum of the $1$-dimensional subspace generated by $v^{i}$, the $1$-dimensional subspace generated by $e^{i}$, and the $2$-dimensional subspace orthogonal to both of them. Hence, the metric tensor can be decomposed as
\begin{equation}
g_{ij} = v_{i}v_{j} + h_{ij},\qquad h_{ij} = - e_{i}e_{j} + N_{ij},
\end{equation}
where $h_{ij}$ is the induced metric on the $3$-dimensional subspace orthogonal to $v^{i}$, and $N_{ij}$ is the induced metric on the $2$-dimensional subspace orthogonal to both $v^{i}$ and $e^{i}$. These tensors satisfy the following conditions
\begin{equation}
h_{ij}v^{j}=0,\quad h_{ij}h^{j}{}_h=h_{ih},\quad h^{i}{}_{i}=3
\end{equation}
\begin{equation}
N_{ij}v^{j}=0,\quad N_{ij}e^{j}=0,\quad N_{ij}N^{j}{}_h=N_{ih},\quad N^{i}{}_{i}=2.
\end{equation}

In LRS spacetimes, we can identify $e^{i}$ as the local axis of rotational symmetry. Accordingly, observations must be invariant under rotations around $e^{i}$, and every physical tensor must have vanishing components along the $2$-dimensional subspace orthogonal to $v^{i}$ and $e^{i}$.

In this general setting, we denote by
\begin{equation}\label{eq:def_time_derivative}
\dot{f} = v^{c}\nabla_{c}f,
\end{equation}
and 
\begin{equation}\label{eq:def_hat_derivative}
\hat{f} = e^{c}\nabla_{c}f
\end{equation}
the derivatives of any scalar function $f$ respectively along $v^{i}$ and $e^{i}$. The covariant derivatives of the vector fields $v^{i}$ and $e^{i}$ are expressed as
\begin{align}
\label{eq:covariant_derivative_v}
\nabla_{i}v_{j}&=\Sigma\Big(e_{i}e_{j}+\frac{1}{2}N_{ij}\Big)+\frac{1}{3}\Theta\big(N_{ij}-e_{i}e_{j}\big)-Av_{i}e_{j}+\Omega\,\varepsilon_{ij}, \\[6pt]
\label{eq:covariant_derivative_e}
\nabla_{i}e_{j}&=\frac{1}{2}\phi\,N_{ij}+\xi\,\varepsilon_{ij} - A v_{i}v_{j}+\Big(\Sigma-\frac{1}{3}\Theta\Big)e_{i}v_{j},
\end{align}
where 
\begin{equation}\label{kinematical_quantities scalars}
\begin{gathered}
A = e^{a}v^{i}\nabla_{i}v_{a} \qquad \Theta=\nabla_{a}v^{a} \qquad \phi=N^{ab}\nabla_{a}e_{b}
\\[6pt]
\Sigma=\left[\frac{1}{2}h_{a}{}^{c}h_{b}{}^{d}(\nabla_{c} v_{d} + \nabla_{d} v_{c})-\frac{1}{3}h_{d}{}^{c}\nabla_{c} v^{d} h_{ab}\right]e^{a}e^{b}  \\[6pt]
\Omega = \frac{1}{2}\varepsilon^{ba}\nabla_{b}v_{a} \qquad
\xi=\frac{1}{2}\varepsilon^{ba}\nabla_{b}e_{a} 
\end{gathered}
\end{equation}
and, given the Levi-Civita tensor $\varepsilon_{hkij}$, we have defined 
\begin{equation}
    \varepsilon_{kij} = \varepsilon_{hkij}v^{h} \qquad \text{and} \qquad \varepsilon_{kij}e^{k} = \varepsilon_{ij}.
\end{equation}
The matter source is described by an energy-momentum tensor of the form
\begin{equation}\label{eq:1+1+2_energy_momentum}
    T_{ab} = \mu v_{a}v_{b} - p(N_{ab} - e_{a}e_{b}) - Q(e_{a}v_{b} + e_{b}v_{a}) + \frac{1}{2}\Pi(N_{ab} + 2e_{a}e_{b})
\end{equation}
with
\begin{equation}
\begin{aligned}
    \mu &=T_{ab}v^{a}v^{b} &  p &=-\frac{1}{3}T_{ab}\left(N^{ab}-e^{a} e^{b}\right) \\[6pt] 
    Q &= T_{ab}e^{a}v^{b} & \Pi &=\frac{1}{3}T_{ab}\left(N^{ab}+2e^{a} e^{b}\right) 
\end{aligned}
\end{equation}
Also, we need to introduce the scalar components of the electric and magnetic parts of the Weyl tensor, respectively denoted by $\mathcal{E}$ and $\mathcal{H}$. 

After that, an LRS spacetime results to be completely characterized by the following set of scalar variables
\begin{equation}\label{LRS_scalars}
\{A,\,\Theta,\,\Sigma,\,\Omega,\,\phi,\,\xi,\,\mathcal{E},\,\mathcal{H},\,\mu,\,p,\,Q,\,\Pi\} \,.
\end{equation}
The covariant equations for the unknowns \eqref{LRS_scalars} in the spinor signature $(+,\,-,\,-,\,-)$ have been deduced in \cite{Vignolo:2025ppo} and are here omitted for the sake of brevity.

\section{Polar formalism of the Dirac field}\label{sec:polar}

The covariant approaches to the Dirac field developed in \cite{Vignolo:2025ppo,Vignolo:2026qfn} rely on the use of the so-called polar formalism. Referring the reader to \cite{Fabbri:2020ypd,Fabbri:2023dgv,Fabbri_2024,Fabbri_2025} for further details, we briefly recall that every regular spinor field $\psi$, as well as its conjugate $\bar\psi = \psi^{\dagger}\gamma^{0}$, admit polar representation of the form
\begin{equation}
\psi=\sqrt{\tfrac{\rho}{2}}\,e^{-\frac{i}{2}\beta\gamma^5}L^{-1}
\begin{pmatrix}1\\0\\1\\0\end{pmatrix},
\end{equation}
where $L$ is a complex matrix with the structure of a complex Lorentz transformation, the scalar $\rho$ and the pseudo-scalar $\beta$ are the modulus and the chiral angle, and $\gamma^{5} = i \gamma^{0}\gamma^{1}\gamma^{2}\gamma^{3}$ is the parity-odd matrix, with $\gamma^{i}$, $i = 0,...,3$, a given set of Clifford matrices.

In the polar formalism, the spinor bilinears can be expressed as
\begin{equation}
\begin{aligned}
i\bar\psi\gamma^{5}\psi&=\rho\sin\beta,& \bar\psi\psi&=\rho\cos\beta, \\[6pt]
\bar\psi\gamma^{a}\gamma^{5}\psi&=\rho s^{a}, & \bar\psi\gamma^{a}\psi&=\rho u^{a},
\end{aligned}
\end{equation}
in terms of the modulus, the chiral angle, and two vector fields $u^{a}$ and $s^{a}$ satisfying the orthonormality conditions
\begin{equation}
u^{a}u_{a} = -s^{a}s_{a}=1, \qquad u^{a}s_{a}=0.
\end{equation}
It is easy to show that in an LRS spacetime, the two pairs $\{v^{i},e^{i}\}$ and $\{u^{i},s^{i}\}$ must be coplanar \cite{Vignolo:2026qfn}. In view of the orthonormality conditions, the pairs $\{v^{i},e^{i}\}$ and $\{u^{i},s^{i}\}$ are related by
\begin{equation}
\begin{aligned}
    v^{i} &= \cosh{\eta}\,u^{i}+\sinh{\eta}\,s^{i}, \\
    e^{i} &= \sinh{\eta}\,u^{i}+\cosh{\eta}\,s^{i} 
\end{aligned}
\end{equation}
or, equivalently,
\begin{equation}
\begin{aligned}
      u^{i} &= \cosh{\eta}\,v^{i}-\sinh{\eta}\,e^{i}, \\
    s^{i} &= -\sinh{\eta}\,v^{i}+\cosh{\eta}\,e^{i},
\end{aligned}\,,
\end{equation}
where $\eta$ is a suitable function of spacetime coordinates. The energy-momentum tensor of the Dirac field
\begin{equation}\label{eq:Dirac_energy_tensor}
T^{ab}=
\frac{i}{8}\left(\bar{\psi}{\gamma}^{a}{\nabla}^{b}\psi
-{\nabla}^{b}\bar{\psi}{\gamma}^{a}\psi
+\bar{\psi}{\gamma}^{b}{\nabla}^{a}\psi
-{\nabla}^{a}\bar{\psi}{\gamma}^{b}\psi\right)
\end{equation}
can be expressed in the form \eqref{eq:1+1+2_energy_momentum} with the associated thermodynamic quantities given by
\begin{align}
\mu&=\frac{1}{2}\rho\left[\left(m\cosh{\eta}\cos{\beta}-\Omega-\frac{1}{2}\hat\beta\right)\cosh{\eta}-\frac{1}{2}\dot{\beta}\sinh{\eta}\right],\\
p&=-\frac{1}{12}\rho\left[\left(2\Omega+\hat\beta\right)\cosh{\eta}+\left(\dot{\beta}-2m\sinh{\eta}\cos{\beta}\right)\sinh{\eta}\right],\\
Q&=
\begin{aligned}[t]
&-\frac{1}{4}\rho\left[\left(\dot{\beta}+\xi\right)\cosh{\eta}\right.\\
&\left.+\left(\hat\beta+\Omega-2m\cosh{\eta}\cos{\beta}\right)\sinh{\eta}\right],
\end{aligned}\\
\Pi&=
\begin{aligned}[t]
&-\frac{1}{6}\rho\left[\left(\hat\beta-\Omega\right)\cosh{\eta}\right.\\
&\left.+\left(\dot{\beta}+3\xi-2m\sinh{\eta}\cos{\beta}\right)\sinh{\eta}\right],
\end{aligned}
\end{align}
Moreover, the Dirac equations reduce to the pair of differential equations
\begin{equation}
\begin{gathered}
\dot{\ln\rho}-\hat{\eta}+\Theta-2m \sinh{\eta}\sin{\beta}=0, \\[6pt]
\hat{\ln\rho}-\dot{\eta}+\phi - A - 2m \cosh{\eta}\sin\beta=0,
\end{gathered}
\end{equation}
where $m$ is the mass of the spinor field.

\section{LRS III spacetimes and non-perfect spinor fluids}\label{sec:LRS III}

LRS III spacetimes are characterized by the conditions $\Omega=0$ and $\xi \neq 0$. As shown in \cite{Vignolo:2025ppo} and \cite{Vignolo:2026qfn}, this type of spacetimes is not compatible with a perfect spinor fluid. Instead, LRS III spacetimes can be sourced by a spinor field provided that the spinor fluid is non-perfect and the pairs $\{v^{i},e^{i}\}$ and $\{u^{i},s^{i}\}$ do not coincide \cite{Vignolo:2026qfn}. In the resulting setting, the condition $Q=0$ must also hold, together with $\phi=A=0$ and $\hat{f}=0$ for any covariantly defined scalar function $f$. The corresponding covariant equations take the form:
\begin{align}
\label{eq:LRS III_1}
\dot{\xi}&=-\frac{1}{3}\xi\left(\Theta+6\Sigma\right)\,,\\[6pt]
\dot{\Theta}&=-\frac{1}{3}\Theta^{2}-\frac{3}{2}\Sigma^{2}-\frac{1}{4}m\rho\cos{\beta}-\frac{1}{4}\rho\xi\sinh{\eta}\,,\\[6pt]
\dot{\Sigma}&=-\frac{1}{2}\Sigma^{2}-\Theta\Sigma+\frac{2}{9}\Theta^{2}+2\xi^{2}-\frac{1}{3}m\rho\cos{\beta}+\frac{1}{6}\rho\xi\sinh{\eta}\,,\\[6pt]
\dot{\rho}&=\rho\left(2m\sinh{\eta}\sin{\beta}-\Theta\right)\,,\\[6pt]
\dot{\eta}&=-2m\cosh{\eta}\sin{\beta}\,,\\[6pt]
\label{eq:LRS III_6}
\dot{\beta}&=2m\sinh{\eta}\cos{\beta}-\xi\,.
\end{align}
Equations \eqref{eq:LRS III_1}-\eqref{eq:LRS III_6} form a system of six differential equations for the six variables $\{\xi,\Theta,\Sigma,\rho,\eta,\beta\}$. The remaining variables, instead, can be expressed in terms of the first six as
\begin{align}
\mu &= \frac{1}{4} \rho\left(2m\cos{\beta}+\xi\sinh{\eta}\right)\,,\\[6pt]
p &= \frac{1}{12}\rho\,\xi\,\sinh{\eta}\,,\\[6pt]
\Pi &= -\frac{1}{3}\rho\,\xi\,\sinh{\eta}\,,\\[6pt]
\mathcal{E} &= \frac{1}{2}\Pi + \frac{2}{3}\mu + \left( \Sigma - \frac{1}{3}\Theta \right)\left( \Sigma + \frac{2}{3}\Theta \right) - 2\xi^{2}\,,\\[6pt]
\mathcal{H} &= -3\xi\,\Sigma\,.
\end{align}
The covariant equations are completed by the Hamiltonian constraint
\begin{equation}\label{eq:LRS III_12}
    -\frac{3}{2}R_{3} + \Theta^{2} - 3\mu - \frac{9}{4}\Sigma^{2} = 0
\end{equation}
where $R_{3}$ is the $3$-curvature of the $3$-dimensional submanifolds orthogonal to $v$. Solving the system \eqref{eq:LRS III_1}-\eqref{eq:LRS III_6} analytically is a very difficult task (some numerical solutions are given in \cite{Vignolo:2026qfn}). However, the evolution and stability of the solutions of Eqs. \eqref{eq:LRS III_1}-\eqref{eq:LRS III_6}, can be investigated qualitatively using the DSA \cite{perko2012differential, Bahamonde:2017ize}. This is precisely the aim of the present work. To proceed with the DSA, we introduce the parameter
\begin{equation}
    \mathbb{B} = \tan{\frac{\beta}{2}}.
\end{equation}
allowing us to express the sine and cosine of $\beta$ as follows
\begin{equation}
    \sin{\beta} = \frac{2\mathbb{B}}{1 + \mathbb{B}^{2}} \,, \qquad \cos{\beta} = \frac{1 - \mathbb{B}^{2}}{1 + \mathbb{B}^{2}}.
\end{equation}
By eliminating the trigonometric dependence on the chiral angle, this substitution significantly simplifies the dynamical analysis.

\section{Dynamical systems analysis}\label{sec:DSA}

The first step in implementing the DSA is to define suitable dimensionless dynamical variables. Rather than using the usual variables that are normalized compared to the expansion $\Theta$, we are normalizing by the quantity $\mathbb{C} = \sqrt{\Theta^{2} + \rho^{2}}$. This choice provides a more comprehensive analysis of the phase space, as it also lets us determine the qualitative value of the expansion at the fixed points. Following the usual procedure, from the Hamiltonian constraint \eqref{eq:LRS III_12}, we introduce the dynamical variables
\begin{equation}\label{eq:dyn_variables}
\begin{aligned}
    x_{\Theta} &= \frac{\Theta}{\mathbb{C}}\,, & x_{\Sigma} &= \frac{3}{2}\frac{\Sigma}{\mathbb{C}}\,, & x_{\rho} &= \frac{3}{2}\frac{\rho}{\mathbb{C}}\,, \\ x_{m} &= \frac{m}{\mathbb{C}}\,, &
    x_{\xi} &= \frac{1}{2}\frac{\xi}{\mathbb{C}}\,, & x_{\mathbb{B}} &= \mathbb{B}\,, \\ x_{\eta} &= \sinh{\eta}\,, & x_{R_{3}} &= \frac{3}{2}\frac{R_{3}}{\mathbb{C}^{2}}.
\end{aligned}
\end{equation}
which have the same functional structure as those commonly used in the literature. We also introduce the dimensionless time variable $\tau$,
\begin{equation}
    \text{d}\tau = \mathbb{C}\,\text{d}t\,.
\end{equation}
The Hamiltonian constraint \eqref{eq:LRS III_12} assumes the form
\begin{equation}\label{eq:fried_dyn_sys}
    x_{R_{3}} = x_{\Theta}^{2}+x_{m}x_{\rho}-\frac{2x_{m}x_{\rho}}{1+x_{\mathbb{B}}^{2}}-x_{\eta}x_{\xi}x_{\rho}-x_{\Sigma}^{2},
\end{equation}
which will give us the value of the curvature at the critical points. From the definition of the variable $\mathbb{C}$, we can also obtain the constraint
\begin{equation}\label{eq:cons_Theta_rho}
    x_{\Theta}^{2} + \frac{4}{9}x_{\rho}^{2} = 1 \quad \Longrightarrow \quad x_{\Theta}\frac{\text{d}x_{\Theta}}{\text{d}\tau} + \frac{4}{9}x_{\rho}\frac{\text{d}x_{\rho}}{\text{d}\tau} = 0, 
\end{equation}
and since $\rho \geq 0$,
\begin{equation}\label{eq:cons_Theta_rho_2}
    x_{\rho} = \frac{3}{2}\sqrt{1 - x_{\Theta}^{2}}
\end{equation}
with $-1 \leq x_{\Theta} \leq 1$. So, by adopting the set of variables $\{x_{\Theta},\, x_{\Sigma},\, x_{m},\, x_{\xi},\, x_{\mathbb{B}},\, x_{\eta}\}$ and using Eqs. \eqref{eq:LRS III_1}-\eqref{eq:LRS III_6}, we obtain the following system of dynamical equations:
\begin{table*}[!t]
    \renewcommand{\arraystretch}{2.0}
    \setlength{\tabcolsep}{4pt}
    \centering
    \begin{tabular}{|c|c|c|c|}
        \hline
        Fixed points & $\{ x_{\Theta},\, x_{\Sigma},\, x_{\rho},\, x_{m},\, x_{\xi},\, x_{\mathbb{B}},\, x_{\eta} \}$ & Stability & $x_{R_{3}}$ \\
        \hline
        \hline
        $P_{1}$ & $\{x_{\Theta},\,-x_{\Theta},\, \frac{3}{2}\sqrt{1-x_{\Theta}^{2}},\, 0,\, 0,\, x_{\mathbb{B}},\, x_{\eta} \}$ & \Gape{\makecell{Attractor for $-1 \leq x_{\Theta}< 0$ \\ Non-hyperbolic for $x_{\Theta} = 0$ \\ Repeller for $0 < x_{\Theta} \leq 1$}} & 0\\
        \hline
        $P_{2}$ & $\{x_{\Theta},\,x_{\Theta},\, \frac{3}{2}\sqrt{1-x_{\Theta}^{2}},\, 0,\, 0,\, x_{\mathbb{B}},\, x_{\eta} \}$ & \Gape{\makecell{Saddle for $-1 \leq x_{\Theta}< 0$ \\ Non-hyperbolic for $x_{\Theta} = 0$ \\ Saddle for $0 < x_{\Theta} \leq 1$ }} & 0 \\
        \hline
        $P_{3}$ & $\{ -1,\, -\frac{1}{2},\, 0,\, 0,\, 0,\, x_{\mathbb{B}},\, x_{\eta} \}$ &  Saddle & $\frac{3}{4}$ \\
        \hline
        $P_{4}$ & $\{1,\, \frac{1}{2},\, 0,\, 0,\, 0,\, x_{\mathbb{B}},\, x_{\eta} \}$ &  Saddle & $\frac{3}{4}$ \\
        \hline
    \end{tabular}
    \caption{Critical points of the system \eqref{eq:dyn_system_1}-\eqref{eq:dyn_system_7} with their stability.}
    \label{tab:fixed_points}
\end{table*}

\begin{align}
\label{eq:dyn_system_1}
\frac{\text{d}x_{\Theta}}{\text{d}\tau} &=
\begin{aligned}[t]
&\frac{x_{\Theta}^{4}}{3}
+\left[x_{m}\left(\frac{1}{6}-\frac{1}{3(1+x_{\mathbb{B}}^{2})}\right)-\frac{1}{3}x_{\eta}\,x_{\xi}\right]x_{\rho}
- \\
&\frac{16\,x_{m}\,x_{\mathbb{B}}\,x_{\eta}\,x_{\Theta}\,x_{\rho}^{2}}{9(1+x_{\mathbb{B}}^{2})} -\frac{2x_{\Sigma}^{2}}{3} +\\
&x_{\Theta}^{2}\left\{-\frac{1}{3}+\left[x_{m}\left(-\frac{1}{6}+\frac{1}{3+3x_{\mathbb{B}}^{2}}\right)+\frac{1}{3}x_{\eta}\,x_{\xi}\right]x_{\rho}+\right.\\
&\left.\frac{4x_{\rho}^{2}}{9}+\frac{2x_{\Sigma}^{2}}{3}\right\}
\end{aligned}\\[10pt]
\label{eq:dyn_system_2}
\frac{\text{d}x_{\Sigma}}{\text{d}\tau} &=
\begin{aligned}[t]
&\frac{x_{\Theta}^{2}}{3}+12x_{\xi}^{2}
+\left[x_{m}\left(\frac{1}{3}-\frac{2}{3(1+x_{\mathbb{B}}^{2})}\right)+\frac{1}{3}x_{\eta}\,x_{\xi}\right]x_{\rho}
+\\
&\frac{1}{3}x_{\Theta}^{3}x_{\Sigma}
-\frac{16\,x_{m}\,x_{\mathbb{B}}\,x_{\eta}\,x_{\rho}^{2}x_{\Sigma}}{9(1+x_{\mathbb{B}}^{2})}-\frac{x_{\Sigma}^{2}}{3} +\\
&x_{\Theta}\left\{\left[-1+\left(x_{m}\left(-\frac{1}{6}+\frac{1}{3+3x_{\mathbb{B}}^{2}}\right)+\frac{1}{3}x_{\eta}\,x_{\xi}\right)x_{\rho}+\right.\right.\\
&\left.\left.\frac{4x_{\rho}^{2}}{9}\right]x_{\Sigma}+\frac{2x_{\Sigma}^{3}}{3}\right\}
\end{aligned}\\[10pt]
\label{eq:dyn_system_4}
\frac{\text{d}x_{m}}{\text{d}\tau} &=
\begin{aligned}[t]
&-\frac{x_{m}^{2}x_{\rho}\left[3(-1+x_{\mathbb{B}}^{2})x_{\Theta}+32x_{\mathbb{B}}x_{\eta}x_{\rho}\right]}{18(1+x_{\mathbb{B}}^{2})}
+\\
&\frac{1}{9}x_{m}x_{\Theta}\left(3x_{\Theta}^{2}+3x_{\eta}x_{\xi}x_{\rho}+4x_{\rho}^{2}+6x_{\Sigma}^{2}\right)
\end{aligned}\\[10pt]
\label{eq:dyn_system_5}
\frac{\text{d}x_{\xi}}{\text{d}\tau} &=
 \begin{aligned}[t]
&\frac{1}{3}x_{\eta}x_{\Theta}x_{\xi}^{2}x_{\rho} +\frac{1}{18(1+x_{\mathbb{B}}^{2})}x_{\xi}\Big\{6(1+x_{\mathbb{B}}^{2})x_{\Theta}^{3}\\
&-32x_{m}x_{\mathbb{B}}x_{\eta}x_{\rho}^{2}-24(1+x_{\mathbb{B}}^{2})x_{\Sigma}\\
&+x_{\Theta}\Big[-6+3x_{m}x_{\rho}+8x_{\rho}^{2}+12x_{\Sigma}^{2} \\
&+x_{\mathbb{B}}^{2}\left(-6-3x_{m}x_{\rho}+8x_{\rho}^{2}+12x_{\Sigma}^{2}\right)\Big]\Big\}
\end{aligned}\\[10pt]
\label{eq:dyn_system_6}
\frac{\text{d}x_{\mathbb{B}}}{\text{d}\tau} &=(1-x_{\mathbb{B}}^{2})x_{m}x_{\eta}-(1+x_{\mathbb{B}}^{2})x_{\xi}\\[10pt]
\label{eq:dyn_system_7}
\frac{\text{d}x_{\eta}}{\text{d}\tau} &= -\frac{4x_{m}x_{\mathbb{B}}\,(1+x_{\eta}^{2})}{1+x_{\mathbb{B}}^{2}}
\end{align}
where $x_{\rho}$ is given by Eq. \eqref{eq:cons_Theta_rho_2}.

It is worth emphasizing that although the vanishing of the variables \eqref{eq:dyn_variables} might suggest the vanishing of the corresponding kinematic and matter quantities, it can also describe a situation in which their evolution is negligible compared to that of $\mathbb{C}$. Consequently, every qualitative label we give to the vanishing of a dynamical variable, e.g., referring to $x_{\rho}=0$ as ``vacuum,'' to $x_{\xi}=0$ as ``LRS II,'' or to $x_{R_{3}}=0$ as ``spatially flat,'' implies both the aforementioned possibilities. The only exception is the case $x_{m}=0$, which will never correspond to $m=0$.

The system \eqref{eq:dyn_system_1}-\eqref{eq:dyn_system_7} exhibits two invariant submanifolds: the invariant submanifold $x_{m} = 0$, for which at least one between $\Theta$ and $\rho$ diverges to infinity; and the invariant submanifold $x_{\xi} = 0$, which instead describes an LRS II spacetime within the dynamical description, filled with a spinor fluid.

\begin{figure}[!t]
\centering
    \begin{subfigure}[t]{0.4\textwidth}
        \includegraphics[width=\linewidth]{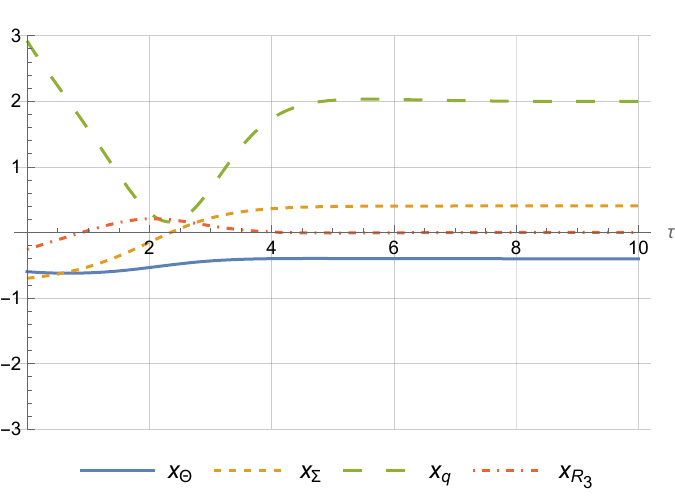}
        \caption{}
        \label{fig:numsol_0_0}
    \end{subfigure}
    \begin{subfigure}[t]{0.4\textwidth}
        \includegraphics[width=\linewidth]{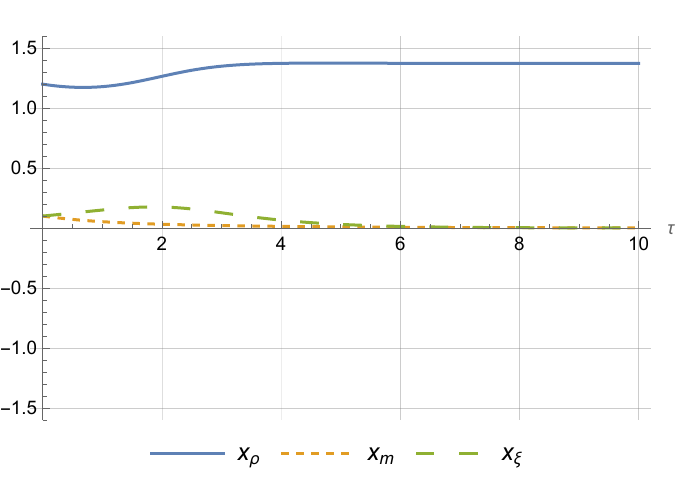} 
        \caption{}
        \label{fig:numsol_0_1}
    \end{subfigure}
    \begin{subfigure}[t]{0.4\textwidth}
        \includegraphics[width=\linewidth]{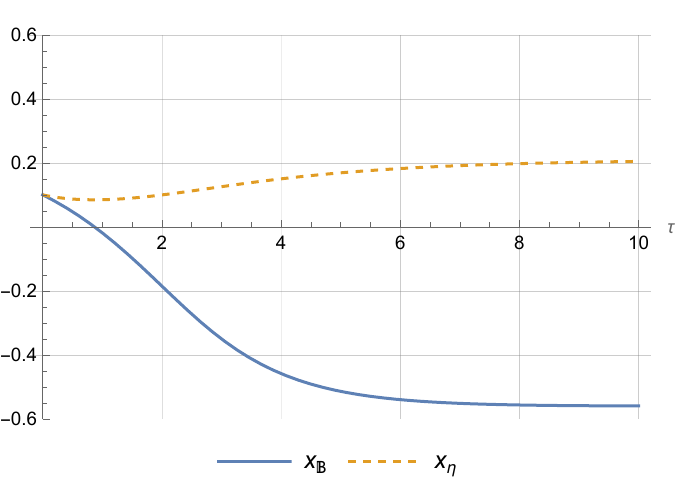}
        \caption{}
        \label{fig:numsol_0_2}
    \end{subfigure}
    \caption{Numerical solutions of the system \eqref{eq:dyn_system_1}-\eqref{eq:dyn_system_7}. In (a) we also plotted the evolution of the deceleration parameter $q$ and $R_{3}$. The initial conditions used are: $x_{\Theta}(0)=-0.6$, $x_{\Sigma}(0)=-0.7$, $x_{m}(0)=x_{\xi}(0)=x_{\mathbb{B}}(0)=x_{\eta}(0)=0.1$.}    
    \label{fig:numsol_0}
\end{figure}

\begin{figure}[!t]
\centering
    \begin{subfigure}[t]{0.4\textwidth}
        \includegraphics[width=\linewidth]{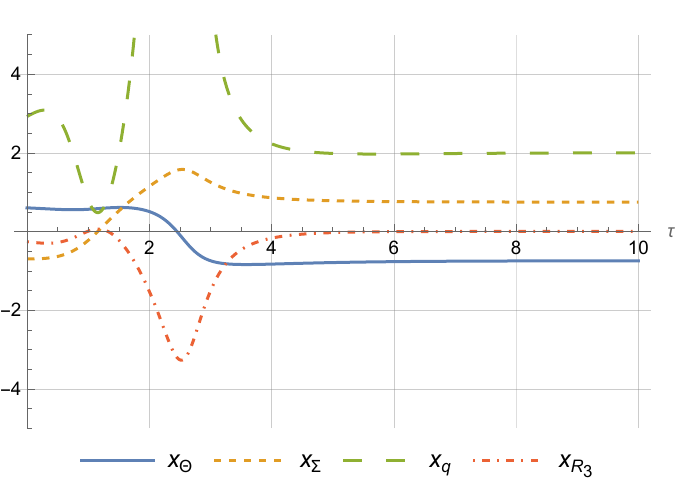}
        \caption{}
        \label{fig:numsol_1_0}
    \end{subfigure}
    \begin{subfigure}[t]{0.4\textwidth}
        \includegraphics[width=\linewidth]{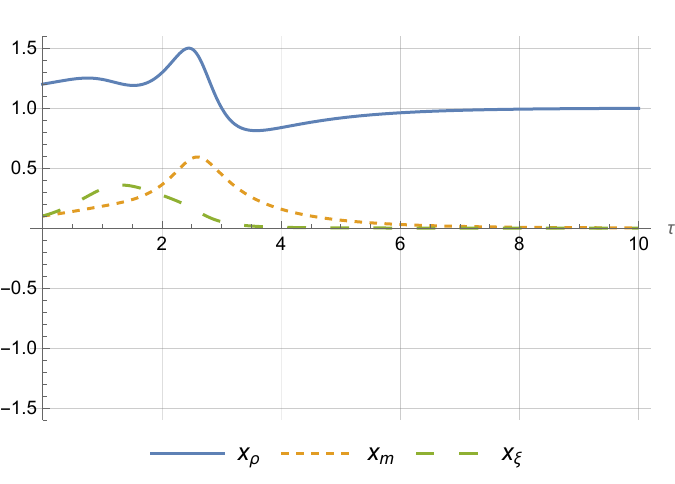} 
        \caption{}
        \label{fig:numsol_1_1}
    \end{subfigure}
    \begin{subfigure}[t]{0.4\textwidth}
        \includegraphics[width=\linewidth]{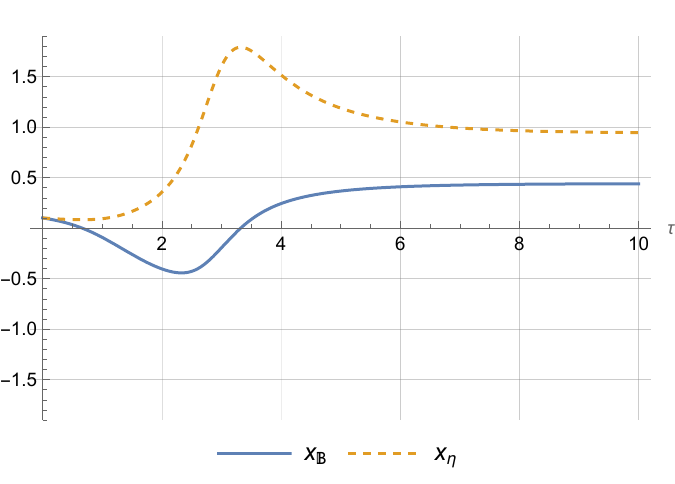}
        \caption{}
        \label{fig:numsol_1_2}
    \end{subfigure}
    \caption{Numerical solutions of the system \eqref{eq:dyn_system_1}-\eqref{eq:dyn_system_7}. In (a) we also plotted the evolution of the deceleration parameter $q$ and $R_{3}$. The initial conditions used are: $x_{\Theta}(0)=0.6$, $x_{\Sigma}(0)=-0.7$, $x_{m}(0)=x_{\xi}(0)=x_{\mathbb{B}}(0)=x_{\eta}(0)=0.1$.}    
    \label{fig:numsol_1}
\end{figure}

The sets of fixed points of the system \eqref{eq:dyn_system_1}-\eqref{eq:dyn_system_7} are listed in Table \ref{tab:fixed_points}. At $x_{\Theta} = 0$, we have that all the eigenvalues vanish, so the corresponding fixed point is non-hyperbolic. Moreover, since all the fixed points lie at the intersection of the invariant submanifolds $x_{m} = 0 $ and $x_{\xi} = 0$, they represent LRS II spacetimes.

The sets of points $P_{1}$ and $P_{2}$ correspond to spatially flat spacetimes filled with a spinor fluid, with the exception of $x_{\Theta} = \pm 1$, which instead represent vacuum spacetimes. However, the points of $P_{1}$ are attractors for $-1 \leq x_{\Theta} < 0$ and repellers for $0 < x_{\Theta} \leq 1$, describing contracting and expanding Bianchi I cosmologies \cite{Ellis2006}, respectively. On the other hand, the points of $P_{2}$ are saddle points of the phase space. At $x_{\Theta} = 0$, $P_{1}$ and $P_{2}$ coincide at the same non-hyperbolic point. Unlike $P_{1}$ and $P_{2}$, which reduce to vacuum spacetimes only at the boundary values $x_{\Theta} = \pm 1$, the points of $P_{3}$ and $P_{4}$ represent vacuum universes and describe Bianchi III cosmologies since the $3$-curvature $R_{3}$ is positive\footnote{Because of the signature $\{+,-,-,-\}$ and Eq. \eqref{eq:LRS III_12}, open universes are represented by a positive value of the curvature $R_{3}$.}. Dynamically, they are saddle points. Overall, the phase-space dynamics is characterized by trajectories that depart from the repelling branch of $P_{1}$ ($x_{\Theta} > 0$) and approach the attracting branch of $P_{1}$ ($x_{\Theta} < 0$), undergoing transitional phases due to $P_{2}$, $P_{3}$, and $P_{4}$.

In general, it is not possible to give a graphical representation of the streamlines for the entire system because of the large number of dynamical variables. Therefore, to show some examples of how the system actually evolves, in Fig. \ref{fig:numsol_0}, we give a numerical solution for initial conditions close to the set of points $P_{2}$. In the plots, the final plateau corresponds to the attractive part of the points $P_{1}$. In Fig. \ref{fig:numsol_0_0}, we also show the behavior of the deceleration parameter $q$,
\begin{equation}\label{eq:dec_par}
\begin{gathered}
    q = -1 - 3\frac{\dot\Theta(t)}{\Theta(t)^{2}} \quad \Longrightarrow \quad
    q=x_{q} \quad \Longrightarrow \\
    x_{q} = \left[ \left( -\frac{1}{2} +
    \frac{1}{1 + x_{\mathbb{B}}^{2}}\right)x_{m} + x_{\eta}x_{\xi} \right]
    \frac{x_{\rho}}{x_{\Theta}^{2}} + 2\frac{x_{\Sigma}^{2}}{x_{\Theta}^{2}},
\end{gathered}
\end{equation}
which after reaching a minimum at the vanishing of $x_{\Sigma}$, tends to the constant value of $2$, as expected by looking at equation \eqref{eq:dec_par}, since at $P_{1}$ we have $x_{m}=x_{\xi}=0$ and $x_{\Sigma}=-x_{\Theta}$. Moreover, $q$ remains positive at all times, while $x_{\Theta}$ is negative along the whole trajectory, so the universe undergoes a contracting evolution and $x_{q}$ in Eq. \eqref{eq:dec_par} remains finite. For completeness, we also plot $x_{R_{3}}$ that, after a first phase in which it is negative, becomes positive and tends to zero, consistently with the fact that $P_{1}$ represents a spatially flat universe.
As an additional example, we also give a numerical solution for initial conditions close to the repelling branch of $P_{1}$, Fig. \ref{fig:numsol_1}. Unlike the trajectory shown in Fig. \ref{fig:numsol_0}, here the variable $x_{\Theta}$ changes sign. This is directly reflected in the behavior of $x_{q}$, which diverges as $x_{\Theta}$ crosses zero. This divergence of the parameter $q$ is typical in models admitting a transition from contraction to expansion, or vice versa, and does not represent any singular behavior in the dynamical variables. Indeed, as can be seen from Figs. \ref{fig:numsol_1_1}-\ref{fig:numsol_1_2}, the other dynamical quantities remain finite. After crossing this point, the trajectory tends to the same attracting branch of $P_{1}$ reached in Fig. \ref{fig:numsol_0}.

\section{Conclusions}\label{sec:conclusions}

In this work, we have performed a dynamical systems analysis of LRS III spacetimes sourced by a self-gravitating, non-perfect spinor fluid within the $1+1+2$ covariant formalism. While LRS II geometries have been extensively employed in cosmology and astrophysics, LRS III spacetimes have received comparatively little attention and, to the best of our knowledge, this work provides the first phase-space analysis of Einstein-Dirac LRS III cosmologies.

The system \eqref{eq:dyn_system_1}-\eqref{eq:dyn_system_7} admits four sets of equilibrium points, all located in the invariant submanifolds $x_{m}=0$ and $x_{\xi}=0$, corresponding to LRS II spacetimes. $P_{1}$ and $P_{2}$ represent spatially flat Bianchi I cosmologies sourced by a spinor fluid, whereas $P_{3}$ and $P_{4}$ correspond to vacuum Bianchi III solutions. The stability analysis shows that only the points of $P_{1}$ can act as attractors for $-1 \leq x_{\Theta}<0$, while they become repellers in the expanding branch $0 < x_{\Theta} \leq 1$. The remaining equilibrium points are all saddle points and therefore govern only transient phases of the evolution.

Direct numerical integrations of the complete dynamical system support the stability results, since we see that the cosmological evolution undergoes transient phases before approaching the attracting Bianchi I branch of $P_{1}$, consistently with the saddle character of the remaining set of fixed points. In this asymptotic regime, the contributions due to the Dirac mass and the spacetime twist become dynamically irrelevant. These results provide a dynamical systems explanation of the numerical behavior previously reported in \cite{Vignolo:2026qfn}, demonstrating that the asymptotic approach to an LRS II geometry emerges from the dynamics of the Einstein-Dirac system rather than being a property of the particular solutions considered there.

An important remark concerns the nomenclature that we have used throughout this work. For our analysis, we have used the pair of vector fields $\{v^{i},e^{i}\}$ which do not necessarily correspond to a description of the spacetime evolution made by an observer comoving with the fermion field. This implies that, in the classification of solutions associated with the fixed points, names like Bianchi I and Bianchi III refer to these observers. Since these names are normally used for observers comoving with the matter fluid and the corresponding geometries are fundamentally different, we have adopted a slight abuse of notation. Nevertheless, we feel this notation is justified, as it helps characterize the fixed points from the perspective of the observers we chose for this work.

The framework developed here can be naturally extended beyond General Relativity. A particularly promising direction is provided by the Riemann-Cartan geometry, where torsion couples directly to the intrinsic spin of the Dirac field. Extending the present dynamical systems analysis to Einstein-Cartan gravity would make it possible to determine whether the contracting Bianchi I attractor persists in the presence of torsion and whether qualitatively new asymptotic regimes arise. More generally, the covariant formalism employed here offers a suitable starting point for investigating the global dynamics of self-gravitating spinor fields in a broad class of modified theories of gravity.

\section*{Acknowledgements}
This work has been carried out in the framework of activities of the INFN Research Project QGSKY.


\bibliographystyle{elsarticle-num} 
\bibliography{bibliography}






\end{document}